\documentstyle[preprint,aps,epsfig]{revtex}

\title{Polarization observables in $p-d$ scattering below 30 MeV}
\author{A. Kievsky$^1$, M. Viviani$^1$ and S. Rosati$^{1,2}$}
\address{Istituto Nazionale di Fisica Nucleare, Via Buonarroti 2, 56100 Pisa, 
Italy} 
\address{Dipartimento di Fisica, Universita' di Pisa, Via Buonarroti 2,
56100 Pisa, Italy}

\def\x{{\bf x}}
\def\y{{\bf y}}
\def\r{{\bf r}}
\def\a{{\alpha}}
\def\r{{\rho}}
\def\aa{{\alpha\alpha'}}
\def\kk{{kk'}}

\def\be{\begin{equation}}
\def\ee{\end{equation}}
\def\bea{\begin{eqnarray}}
\def\eea{\end{eqnarray}}
\def\ra{\rightarrow}
\def\htm{{\hbar^2\over M_N}}
\begin{document}

\maketitle

\begin{abstract}
Differential and total breakup cross sections as well as vector and 
tensor analyzing powers for p-d scattering are studied for
energies above the deuteron breakup threshold up to $E_{lab}=28 $ MeV.
The p-d scattering wave function is expanded in terms of the
correlated hyperspherical harmonic basis and the elastic $S$-matrix
is obtained using the Kohn variational principle in its
complex form. The effects of the Coulomb interaction, which are
expected to be important in this energy range, have been rigorously
taken into account.  
The Argonne AV18 interaction and the Urbana URIX three-nucleon potential 
have been used to perform a comparison to the available experimental data.
\end{abstract}

\newpage

\section{Introduction}

In ref.~\cite{KRV99} the authors recently presented an application
of the Kohn Variational Principle (KVP) in its complex form to calculate
the elastic observables in p-d scattering for energies above the
deuteron breakup threshold (DBT). The KVP was implemented to describe
continuum states of three outgoing particles including
the distortion due to the Coulomb interaction in the asymptotic region.
Only two energies were considered, $E_{lab}=5$ and $10$ MeV.
The validity of the KVP for the elastic 
$S$-matrix describing the $2\rightarrow 2$ process in p-d scattering
for energies above the DBT
has been extensively discussed in ref.~\cite{VKR00}. 

In the present paper the analysis of the elastic p-d reaction is extended
up to $E_{lab}=28$ MeV, covering the region where Coulomb effects are 
expected to be important. The large amount of accurate experimental data 
allows for interesting comparisons. It should be noted that, at present, 
the analysis of the polarization data at energies above the DBT has been
done mainly by comparing n-d calculations to p-d data~\cite{report,witala}. 
Differential cross section and vector analyzing power data exist for both
n-d and p-d scattering, allowing for an estimate of the Coulomb effects. 
Conversely, no n-d data are available for the deuteron analyzing powers
$iT_{11}$, $T_{20}$, $T_{21}$, $T_{22}$. These quantities are evaluated
from experiments using a polarized deuteron beam on unpolarized proton 
targets. The inverse experiment of unpolarized proton or neutron beams on 
a polarized deuteron target seems to be extremely difficult at low energies
and has not yet been done.
 
Experiments using charged particles are certainly easier to perform
and show smaller error bars than those using a neutral beam. On the other
hand, the theoretical description of collisions with more than one 
charged particle in the final state has represented a difficult problem for
many years. In a recent work a complete solution of the 
reaction e$^-$ + H $\rightarrow$ H$^+$ + e$^-$ + e$^-$ has been
obtained by Rescigno {\sl et al.}~\cite{res99} by transforming the 
Schr\"odinger equation using the so-called exterior complex scaling and 
making use of supercomputers to solve the associated equations numerically. 
This was the first complete solution of a three-body collision with all the
charged particles moving away from each other in the final state.
Regarding the p-d reaction, different techniques have been applied so far. 
The Faddeev equations in momentum space have been adapted to take into account
the long-range Coulomb interaction using the screening and
renormalization approach~\cite{alt}. 
Recently, a detailed comparison between the solutions of the
Faddeev equations in configuration space and the KVP has been performed,
though restricted to energies below the DBT~\cite{kp01}.
In the present work we turn our attention to describing p-d
elastic observables above the DBT. In this case the application
of the KVP is feasible and the calculation
of the elastic $S$-matrix does not require large computational devices. 

The study of the three-nucleon (3N) continuum provides important
information about the capability
of modern NN potentials to describe the three-nucleon
dynamics. At present, a few realistic NN potentials are available that
reproduce a large set of two-nucleon (2N) data with $\chi^2 \approx 1$
(per datum). 
They are substantially equivalent in reproducing
all the details of the NN scattering, but in the description of
nuclear systems with $A>2$ differences appear. 
In addition, the three-nucleon system is the simplest one
in which three-nucleon force (3NF) effects can be studied.
The first signal for the necessity of a 3NF comes from 
the underbinding of the triton when only NN forces are used. 
Widely used 3NF models are based on the exchange of two 
pions with an intermediate $\Delta$ excitation.  In general these
models include a certain numbers of parameters which are not precisely
determined by theory, so some of them can be taken as free parameters
in order to reproduce, for example, the triton or $^3$He binding energy. 
As a consequence,
other observables which scale with the three-nucleon binding energy
improve as well. Examples are the bound state r.m.s radii and the zero energy
total cross section in n-d and n-$^3$H scattering. On the contrary,
vector and tensor N-d analyzing powers do not present such a scaling.

Accurate measurements of p-d observables below the DBT, have been
reported recently~\cite{shimizu,knutson,brune}. A comparison of the 
theoretical predictions to these data shows an underprediction of the deuteron
vector analyzing power $iT_{11}$ by $\approx 30$\%~\cite{KRTV96}. A 
similar discrepancy had been observed earlier in the neutron analyzing power
$A_y$, a problem which is usually known as the $A_y$ puzzle~\cite{puzzle}.
As the energy increases the observed discrepancies in $A_y$ and $iT_{11}$
reduce and tend to disappear, though not completely, above $30$ 
MeV~\cite{report}. Accordingly, the study of these observables over the 
energy region considered here is important for understanding such a
behavior.

Accurate 3N and 4N scattering wave functions are necessary for calculating
a number of nuclear reactions. The technique used in the present work 
is based on the expansion of the
wave function in terms of Jastrow type Correlated Hyperspherical Harmonic
(CHH) basis functions. When the correlation factor reduces to a pair 
correlation function the Pair Correlated Hyperspherical Harmonic (PHH) 
basis is obtained.  The CHH and PHH bases have been used to calculate 
the bound states of the $A=3,4$ nuclei~\cite{KVR93,VKR95}, N-d 
scattering~\cite{KVR94,KVR95}, and p-$^3$He and n-$^3$H 
scattering~\cite{VRK98} at energies below the three-body fragmentation. 
Moreover, wave functions obtained through those expansions have 
recently been used to study the radiative capture 
$p+d\rightarrow {}^3{\rm He}+\gamma$ below the DBT~\cite{Viv00} and 
the hep process, namely the weak capture 
$p+{}^3{\rm He}\rightarrow{}^4{\rm He}+e^++\nu_e$ 
at the Gamow peak~\cite{Mar00}. These two reactions have considerable 
astrophysical relevance.
The former is the second reaction in the $pp$ solar chain and has a
prominent role in the evolution of protostars whereas the hep process plays
an important role in the solar neutrino problem. The calculation of the
p-d wave functions above the DBT will provide the input for further
studies of radiative capture and photo and electrodisintegration of
$^3$He. 

In the present paper we present the results obtained for the differential 
and total breakup cross sections, nucleon analyzing
powers $A_y$ and deuteron analyzing powers $iT_{11}$, $T_{20}$,
$T_{21}$ and $T_{22}$ for N-d scattering at different energies. 
The calculations have been done using the two-nucleon AV18 
potential~\cite{av18} with and without the three-nucleon 
URIX force~\cite{urbana}. The results are given at nine different energies
in the range $5$ MeV $\leq E_{lab} \leq 28$ MeV. It has to be noted that,
disregarding small corrections, $E_{c.m.}={2\over 3}E_N$ $({1\over 3}E_d$),
where $E_N$ ($E_d$) is the nucleon (deuteron) incident energy and in the
following we define $E_{lab} \equiv E_N$. The highest energy considered here
is $E_d=56$ MeV ($E_{lab}=28$ MeV) at which the deuteron analyzing
powers are available~\cite{hatanaka84}.  Just above the DBT,
deuteron vector and tensor analyzing powers are available at
$E_{lab}=5$ MeV~\cite{kexp}. For $E_{lab}\leq 18$ MeV 
differential cross sections, proton and deuteron analyzing powers 
have been measured at several energies~\cite{sagara}. In ref.~\cite{bruebler83}
differential cross section as well as vector and tensor observables have 
been measured between $8.5$ MeV $\leq E_{lab}\leq 22.7$ MeV, 
though data for $T_{21}$ are missing at some energies. 

The paper is organized as follows. In Section II the
Kohn variational principle is reviewed. In Section III the
numerical solution of the related differential equations are
compared to previous results. Cross sections and observables are
compared to the data in Section IV, and the conclusions are given 
in the last section.

\section{The Kohn Variational Principle above the deuteron breakup threshold}

In the literature several investigations regarding the validity of the KVP 
above the DBT can be found, starting with the works of Nuttall~\cite{nuttall} 
and Merkuriev~\cite{merkuriev}, where 
the discussion, however, was limited to the n-d reaction. 
The first extensive demonstration of the applicability of the principle 
to the p-d collision has been given in ref.~\cite{VKR00}. The main result
derived in~\cite{VKR00}
is that the effect of the Coulomb interaction can be taken 
into account in such a way that the form of the principle 
remains unchanged when the energy goes from below to above the DBT. 
Below the DBT the
collision matrix is unitary and the problem can be formulated
in terms of the real reactance matrix ($K$--matrix).
Above the DBT the elastic part of the collision matrix
is no longer unitary and the formulation in terms of the
$S$-matrix, the complex form of the KVP, is convenient.
Refering to ref.~\cite{VKR00} for details, a brief description
of the method is given below. The scattering wave function (w.f.) $\Psi$ is 
written as sum  of two terms:
 \begin{equation}
   \Psi=\Psi_C+\Psi_A \ .
\label{eq:psi}
 \end{equation}
The first term, $\Psi_C$, describes the
system when the three--nucleons are close to each other. For large
interparticle separations and energies below the 
DBT it goes to zero, whereas for higher energies it must
reproduce a three outgoing particle state. It
is written as a sum of three Faddeev--like amplitudes
corresponding to the three cyclic permutations of the particle indices
1, 2 , 3. Each amplitude $\Psi_C(\x_i,\y_i)$, where $\x_i,\y_i$ are
the Jacobi coordinates corresponding to the $i$-th permutation, has
total angular momentum $JJ_z$ and total isospin $TT_z$ and is
decomposed into channels using $LS$ coupling, namely
\begin{eqnarray}
     \Psi_C(\x_i,\y_i) &=& \sum_{\alpha=1}^{N_c} \phi_\alpha(x_i,y_i) 
     {\cal Y}_\alpha (jk,i)  \\
     {\cal Y}_\alpha (jk,i) &=&
     \Bigl\{\bigl[ Y_{\ell_\alpha}(\hat x_i)  Y_{L_\alpha}(\hat y_i) 
     \bigr]_{\Lambda_\alpha} \bigl [ s_\alpha^{jk} s_\alpha^i \bigr ]
     _{S_\alpha}
      \Bigr \}_{J J_z} \; \bigl [ t_\alpha^{jk} t_\alpha^i \bigr ]_{T T_z},
\end{eqnarray}
where $x_i,y_i$ are the moduli of the Jacobi coordinates and
${\cal Y}_\alpha$ is the angular-spin-isospin function for each channel.
The maximum number of channels considered in the expansion is $N_c$.
The two-dimensional amplitude $\phi_\alpha$ is expanded in terms of the
PHH basis
\begin{equation}
     \phi_\alpha(x_i,y_i) = \rho^{-5/2} f_\alpha(x_i) 
     \left[ \sum_K u^\alpha_K(\rho) {}^{(2)}P^{\ell_\alpha,L_\alpha}_K(\phi_i)
     \right] \ ,
\label{eq:PHH}
\end{equation}
where the hyperspherical variables, the hyperradius $\rho$ and
the hyperangle $\phi_i$, are defined by the relations
$x_i=\rho\cos{\phi}_i$ and $y_i=\rho\sin{\phi}_i$. The factor
${}^{(2)}P^{\ell,L}_K(\phi)$ is a hyperspherical polynomial and
$f_\alpha(x_i)$ is a pair correlation function introduced
to accelerate the convergence of the expansion. For small values
of the interparticle distance $f_\alpha(x_i)$ is regulated by the
NN interaction whereas for large separations the correlation function
is chosen to satisfy $f_\alpha(x_i)\rightarrow 1$~\cite{KVR93}.

The second term, $\Psi_A$, in the variational wave function of
eq.(\ref{eq:psi}) 
describes the asymptotic  motion of a deuteron relative to the third
nucleon. It can also be written  as a sum
of three amplitudes with the generic one having the form
\begin{equation}
   \Omega^\lambda_{LSJ}(\x_i,\y_i) = \sum_{l_\a=0,2} w_{l_\a}(x_i)
       {\cal R}^\lambda_L (y_i) 
       \left\{\left[ [Y_{l_\a}({\hat x}_i) s_\a^{jk}]_1 s^i \right]_S
        Y_L({\hat y}_i) \right\}_{JJ_z}
       [t_\a^{jk}t^i]_{TT_z}\ , \label{eq:omega}
\end{equation}
where $w_{l_\a}(x_i)$ is the deuteron w.f. component in the state $l_\a =0,2$.
In addition, $s_\a^{jk}=1,t_\a^{jk}=0$ and $L$ is the relative angular momentum 
of the deuteron and the incident nucleon. The superscript $\lambda$ indicates 
the regular ($\lambda\equiv R$) or the irregular ($\lambda\equiv I$)
solution. In the $p-d$ ($n-d$) case, the functions
${\cal R}^\lambda$ are related to
the regular or irregular  Coulomb (spherical Bessel) functions.
The functions $\Omega^\lambda$ can be combined to form a general
asymptotic state 
\begin{equation}
\Omega^+_{LSJ}(\x_i,\y_i) =  \Omega^0_{LSJ}(\x_i,\y_i)+
 \sum_{L'S'}{}^J{\cal L}^{SS'}_{LL'}\Omega^1_{L'S'J}(\x_i,\y_i)  \ ,
\end{equation}
where
\begin{eqnarray}
\Omega^0_{LSJ}(\x_i,\y_i) =& u_{00}\Omega^R_{LSJ}(\x_i,\y_i)+
                            u_{01}\Omega^I_{LSJ}(\x_i,\y_i) \ , \\
\Omega^1_{LSJ}(\x_i,\y_i) =& u_{10}\Omega^R_{LSJ}(\x_i,\y_i)+
                            u_{11}\Omega^I_{LSJ}(\x_i,\y_i)  \ .
\end{eqnarray}
The matrix elements $u_{ij}$ can be selected according to the
four different choices of the matrix ${\cal L}=$ $K$-matrix, 
$K^{-1}$-matrix, $S$-matrix or $T$-matrix. A general
three-nucleon scattering w.f. for an incident 
state with relative angular momentum $L$, spin $S$ and total angular momentum
$J$ is
\begin{equation}
\Psi^+_{LSJ}=\sum_{i=1,3}\left[ \Psi_C(\x_i,\y_i)+\Omega^+_{LSJ}(\x_i,\y_i)
             \right] \ ,
\end{equation}
and its complex conjugate is $\Psi^-_{LSJ}$. A variational estimate of the
trial parameters in the w.f. $\Psi^+_{LSJ}$ can
be obtained by requiring, in accordance with
the generalized KVP, that the functional 
\begin{equation}
[{}^J{\cal L}^{SS'}_{LL'}]= {}^J{\cal L}^{SS'}_{LL'}-{2\over {\rm det}(u)}
\langle\Psi^-_{LSJ}|H-E|\Psi^+_{L'S'J}\rangle \ ,
\label{eq:kohn}
\end{equation}
be stationary. Below the DBT due to the unitarity of the
$S$-matrix, the four forms for the ${\cal L}$-matrix are equivalent. 
However, it was shown that when the
complex form of the principle is used, there is a considerable
reduction of numerical instabilities~\cite{lucc}. Applications of
the complex KVP for N-d scattering (below the DBT)
can be found in ref.~\cite{kiev97}. 
Above the DBT it is convenient to formulate the variational principle 
in terms of the $S$--matrix. Accordingly, we get the following functional:
\begin{equation}
[{}^J{S}^{SS'}_{LL'}]= {}^J{S}^{SS'}_{LL'}+{i}
\langle\Psi^-_{LSJ}|H-E|\Psi^+_{L'S'J}\rangle \ .
\label{eq:ckohn}
\end{equation}

The variation of the functional with respect to the hyperradial 
functions $u^\alpha_K(\rho)$
leads to the following set of coupled equations (hereafter named SE1):
\be
  \sum_{\alpha',k'}
  \Bigl[ A^\aa_\kk (\r ){d^2\over d\r^2}+ B^\aa_\kk (\r ){d\over d\r}
        + C^\aa_\kk (\r )+
     {M_N\over\hbar^2} E\; N^\aa_\kk (\r )\Bigr ]
        u^{\alpha'}_{k'}(\r)= D^\lambda_{\alpha k}(\rho) \ .
   \label{eq:siste}
\ee
For each asymptotic state $^{(2S+1)}L_J$  two different inhomogeneous terms
are constructed corresponding to the asymptotic $\Omega^\lambda_{LSJ}$ 
functions with $\lambda\equiv 0,1$. Accordingly, two sets of solutions
are obtained and
combined to minimize the functional (\ref{eq:ckohn}) with respect to
the $S$-matrix elements. This is the first order solution, the second order 
estimate of the $S$-matrix is obtained after replacing the first order 
solution in eq.(\ref{eq:ckohn})~\cite{KVR94,kiev97}. 

In order to solve the system SE1 appropriate
boundary conditions must be specified for the hyperradial functions. 
For energies below the  DBT  they go to zero when $\rho\rightarrow\infty$, 
whereas
above the DBT energy they asymptotically describe the breakup configuration. 
The boundary conditions to be applied in this case have
been discussed in refs.~\cite{VKR00,KVR97} and are briefly illustrated below.
To simplify the notation let us label the basis
elements with the index $\mu\equiv[\a,K]$,
and introduce the following completely antisymmetric correlated
spin-isospin-hyperspherical basis elements
\be\label{eq:bco}
     {\cal P}_\mu(\rho,\Omega)= \sum_{i=1}^3
      f_\a(x_i)\; {}^{(2)}P^{\ell_\alpha,L_\alpha}_K(\phi_i)
      {\cal Y}_\alpha(jk,i) \ , 
\ee
which depend on $\rho$ through the correlation factor and form a
non--orthogonal basis.
In terms of the ${\cal P}_\mu(\rho,\Omega)$ the internal part is written as
\be
   \Psi_C= \rho^{-5/2}\sum_{\mu=1}^{N_m}
    u_{\mu}(\rho) {\cal P}_\mu(\rho,\Omega) \ ,
\ee
with $N_m$ the total number of basis functions considered.
The ``uncorrelated'' basis elements ${\cal P}^0_\mu(\Omega)$
are obtained from eq.~(\ref{eq:bco}) by setting all the correlation
functions $f_\a(x_i)=1$. It is important to note that 
the elements ${\cal P}^0_\mu(\Omega)$ do not form an orthogonal
basis, as has been discussed in ref.~\cite{KMRV97} where the standard
hyperspherical harmonic basis (HH) has been used to calculate the
three--nucleon bound state. Those basis
elements having the same grand--angular quantum number $G_\mu=\ell_\alpha +
L_\alpha + 2K$, the same $\Lambda_\alpha$ and $S_\alpha$,
but belonging to different channels, are not orthogonal to each others. 
Moreover, some of them are linearly dependent. 
In Ref.~\cite{KMRV97} such states have been identified and removed from 
the expansion used to describe the triton bound state.

In the present case, the basis elements
${\cal P}_\mu(\rho\ra\infty,\Omega)$ reduce to the
uncorrelated ones ${\cal P}^0_\mu(\Omega)$ in the
asymptotic region since $f_\alpha(x)\rightarrow 1$ for large interparticle 
distances. Therefore, it appears useful to
combine the correlated basis~(\ref{eq:bco}) in order to
define a new basis with the property of being orthonormal when
$\rho\rightarrow\infty$.
This can be readily accomplished by noting that the matrix elements
of the norm $N$ behave as
\be
   N_{\mu\mu'}(\rho)= \int d\Omega\;
     {\cal P}_\mu(\rho,\Omega)^\dag
     {\cal P}_{\mu'} (\rho,\Omega) \ra N^{(0)}_{\mu\mu'}
    +{  N^{(3)}_{\mu\mu'}\over \rho^3}+{\cal O}(1/\rho^5)
      \ ,\qquad {\rm for\ }\rho\ra\infty\ ,  \label{eq:n}
\ee
where, in particular,
\be
   N^{(0)}_{\mu\mu'}= \int d\Omega\;
      {\cal P}^0_\mu (\Omega)^\dag
      {\cal P}^0_{\mu'} (\Omega)\ .  \label{eq:n1}
\ee
Let us define a  matrix $U$ such that the matrix $ U^t \; N^{(0)} \; U=
{\cal N}$ is diagonal with diagonal elements ${\cal N}_\mu$
either $1$ or $0$.
The values ${\cal N}_\mu=0$ correspond to states ${\cal P}^0_\mu
(\Omega)$ that depend linearly on others. 
New uncorrelated and correlated bases are defined as:

\be\label{eq:Qor}
   {\cal Q}^0_\mu(\Omega) \equiv \sum_{\mu'=1}^{N_m} U_{\mu'\mu} {\cal
   P}^0_{\mu'}(\Omega)\ ,\qquad
   {\cal Q}_\mu(\rho,\Omega) \equiv \sum_{\mu'=1}^{N_m} U_{\mu'\mu} {\cal
   P}_{\mu'}(\rho,\Omega)\ ,
\ee
The basis functions ${\cal Q}_\mu(\rho,\Omega)$ are
still not orthogonal for any finite values of $\rho$. 
When $\rho\ra\infty$, the elements 
${\cal Q}_\mu(\rho,\Omega)\ra {\cal Q}^0_\mu(\Omega)$. 
Due to the fact that some of the uncorrelated elements ${\cal
P}^0_\mu(\Omega)$ are linearly dependent, some elements
${\cal Q}^0_\mu (\Omega)$ are identically zero. Therefore, some
correlated elements
have the property: ${\cal Q}_\mu(\rho,\Omega) \ra 0$ as $\rho\ra\infty$.
In the following we arrange the new basis in such a way that for values
of the index $\mu\le\overline{N}_m$ the eigenvalues of the
norm are ${\cal N}_\mu=1$ and
for $\overline{N}_m+1 \le \mu \le N_m$ they are ${\cal N}_\mu=0$.

In terms of the new basis, the internal part  $\Psi_C$ is simply
\be\label{eq:Qor2}
   \Psi_C=\rho^{-5/2}
          \sum_{\mu=1}^{N_m} \omega_\mu(\rho) {\cal Q}_\mu (\rho,\Omega)
   \ ,
\ee
where the old set of hyperradial functions is related to the new
set through the transformation $u_\mu=\sum_{\mu'}U_{\mu\mu'}\; \omega_{\mu'}$. 
The variation of the functional~(\ref{eq:kohn}) with respect to the new
hyperradial functions $\omega_\mu(\rho)$, which are now the unknown quantities
entering into the description of the internal part of the w.f. $\Psi_C$,
leads to a set of inhomogeneous second order differential equations
formally equal to SE1, and hereafter called SE2, in which 
each matrix $X\equiv A,B,C,N$ of eq.(\ref{eq:siste}) is substituted by 
$\overline X = U^t X U$ and
the inhomogeneous term $D^\lambda$ by ${\overline D}^\lambda = U^t D^\lambda$.

For $\rho\ra\infty$, neglecting terms going to zero faster
than $\rho^{-2}$, the asymptotic expression of SE2
reduces to the form
\be \label{eq:c0}
   \sum_{\mu'} \biggl\{
   -\htm  \left( {d^2\over d\rho^2} -{{\cal K}_\mu({\cal K}_\mu +1)\over\rho^2}
   + Q^2 \right ){\cal N}_\mu
    \delta_{\mu,\mu'} +
     {2\;Q\; \chi_{\mu\mu'}\over \rho} \;
      +{\cal O}({1\over\rho^3})\biggr\}\omega_{\mu'}(\rho) = 0 \ ,
\ee
where $E=\hbar^2 Q^2/M_N$, ${\cal K}_\mu= G_\mu+3/2$ and the matrix 
$\chi$ is defined as
\be\label{eq:c}
   { \chi}_{\mu\mu'}= \int d\Omega\;
      {\cal Q}^0_{\mu} (\Omega)^\dag
     \; \hat \chi \;
      {\cal Q}^0_{\mu'} (\Omega)
      \ .
\ee
The dimensionless operator $\hat\chi$ originates from the Coulomb interaction
as
\be
   \hat \chi = {M_N\over 2\hbar^2 Q}
  \sum_{i=1}^3 {e^2\over \cos\phi_i} {1+\tau_{j,z} \over 2}
  {1+\tau_{k,z} \over 2} \ .
\label{eq:chi}
\ee
It should be noticed that $\chi_{\mu\mu'}=0$ if $\mu,\mu'>{\overline{N}_m}$.

In practice,
the functions $\omega_\mu(\rho)$ are chosen to be regular at the origin, i.e.
$\omega_\mu(0)=0$ and, in accordance with the equations to be satisfied for
$\rho\ra\infty$, to have the following behavior
($\mu\le\overline{N}_m$)
\be\label{eq:asy2}
  \omega_\mu(\rho) \rightarrow
   - \sum_{\mu'=1}^{\overline{N}_m}
  \left ( e^{-i {\hat \chi} \ln 2 Q\rho} \right)_{\mu\mu'}\;
   b_{\mu'} \; e^{i Q\rho} \ ,
\ee
where $ b_{\mu'}$ are unknown coefficients. This form corresponds
to the asymptotic behavior of three outgoing particles
interacting through the  Coulomb potential~\cite{merkuriev2}. 
In the case of $n-d$ scattering ($\chi\equiv 0$)
the outgoing solutions evolve as outgoing Hankel functions 
$H^{(1)}(Q\rho)$ ($\omega_\mu(\rho)\rightarrow -b_\mu e^{iQ\rho}$). 

For values of the index $\mu > \overline{N}_m$ the eigenvalues of the
norm are ${\cal N}_\mu=0$ and the leading terms
in eq.(\ref{eq:c0}) vanish. So, the asymptotic behavior of these 
$\omega_\mu$ functions is governed by the next order terms. 
A lengthly analysis of the $1/\rho^3$ and $1/\rho^4$ terms for
each matrix $X\equiv A,B,C,N$ shows that these functions
behave as $e^{i(Q^\prime_\mu\rho - \Sigma_\mu{\rm ln}2Q\rho)}$ where the
quantities $Q^\prime_\mu,\Sigma_\mu$ are related to the asymptotic expansion
of the matrices $A,B,C,N$. This asymptotic behavior has been obtained
neglecting all couplings between the $\mu$-th equation 
($\mu > \overline{N}_m$) and all the others. 
If couplings up to $1/\rho^4$ are taken into account the
quantities $Q^\prime,\Sigma$ become matrices and 
we have ($\mu > \overline{N}_m$)
\be\label{eq:asy3}
  \omega_\mu(\rho) \rightarrow
   - \sum_{\mu'=1}^{N_m}
  \left[e^{(iQ'\rho - \Sigma \ln 2 Q\rho)} \right]_{\mu\mu'}\; c_{\mu'}\ ,
\ee
where the $c_{\mu'}$ are unknown coefficients.
Previously we have shown that, for $\mu > \overline{N}_m$, the elements
${\cal Q}_\mu\rightarrow 0$ as $\rho\rightarrow\infty$. The specific
form of the (complex) matrix $\Sigma$ is such that in all cases
$\omega_\mu{\cal Q}_\mu\rightarrow 0$ as $\rho\rightarrow\infty$.
Accordingly, the states with $\mu > \overline{N}_m$ 
do not contribute to the outgoing flux.

In ref.~\cite{KVR97} the set of equations SE2 has been 
solved numerically by choosing a grid of values for the hyperradius
from the origin up to a certain value $\rho_0$.
The differential operators have been substituted by finite differences 
in such a way that SE2 reduces to a set of linear equations 
that can be solved by standard numerical methods. In order to completely
determine the problem, boundary conditions must be imposed at 
$\rho=\rho_0$. To accomplish this, 
eq.(\ref{eq:c0}) has been solved for $\rho > \rho_0$
taking into account coupling terms up to
$\rho^{-4}$ by an expansion of the functions $w_\mu$ in powers of 
$1/\rho$ and verifying the outgoing boundary conditions 
of eqs.(\ref{eq:asy2},\ref{eq:asy3}).
Then, the continuity of the solutions and their first derivatives
has been imposed at the matching radius $\rho_0$. 
The value of $\rho_0$ is not important provided that the asymptotic expression
of SE2 is already reached. This condition is well verified for
values of the matching radius $\rho_0 \gtrsim  80-100$ fm.
However, the functions
$\omega_\mu(\rho)$ show an oscillatory behavior outside the range of
the potential, typically for hyperradial values $\rho>30$ fm. Therefore
a large number of grid points were necessary to obtain stable solutions.
Thus, in ref.~\cite{KVR97} the calculation of N-d scattering states above
the DBT was restricted to a simplified interparticle potential, namely
an $s$-wave interaction. In such a case the number of coupled equations 
to be considered was sufficiently small. When realistic NN interactions 
are considered the number of coupled equations to take into account 
increases considerably.
As a consequence, the dimension of the matrices after the reduction of
derivatives to finite differences can be quite large.
In order to keep the dimension of the matrices low, an alternative method 
of solution in the region $\rho\le\rho_0$
is to expand the hyperradial functions in terms of Laguerre 
polynomials~\cite{KRV99} plus an auxiliary function

\begin{equation} \label{eq:M}
 \omega_\mu(\rho)=\rho^{5/2}\sum_{m=0}^M A^m_{\mu} L^{(5)}_m(z)\exp(-{z\over 2})
 +A^{M+1}_{\mu} \overline \omega_{\mu}(\rho) \ ,
\end{equation}
where $z=\gamma\rho$ and $\gamma$ is a nonlinear parameter.
The linear parameters $A^m_{\mu}$ $(m=0,....,M+1)$ 
are determined by the variational procedure.
The functions defined above are matched to the outgoing solutions of
eq.~(\ref{eq:c0}) at $\rho=\rho_0$. 

The inclusion of the auxiliary functions $\overline \omega_{\mu}(\rho)$ 
defined in eq.(\ref{eq:M}) is useful for reproducing the oscillatory
behavior shown by the hyperradial functions for $\rho\gtrsim 30$ fm. 
Otherwise 
a rather large number $M$ of polynomials should be included in
the expansion. A convenient choice is to take them
as the solutions of a one dimensional
differential equation corresponding to the $\mu$-th equation of
SE2:
\be
       \Bigl[ \overline A_{\mu\mu}(\r ){d^2\over d\r^2}
        + \overline B_{\mu\mu}(\r ){d\over d\r}
        + \overline C_{\mu\mu}(\r )
        +Q^2 \;\overline N_{\mu\mu}(\r )\Bigr ]
        \overline w_{\mu}(\r)= \overline D^\lambda_{\mu}(\rho) \ .
   \label{eq:siste1}
\ee

The functions $\overline \omega_\mu$ are chosen to be
regular at the origin and they are matched to the solutions
of eq.(\ref{eq:c0}) which have been obtained through an
expansion in inverse powers of $\rho$ as has been previously discussed.
For $\mu>\overline{N}_m$ the matching at $\rho_0$ has been done
disregarding the couplings between the different equations
in the region $\rho>\rho_0$, i.e. $\overline\omega_\mu(\rho) \rightarrow
e^{i(Q^\prime_\mu\rho - \Sigma_\mu{\rm ln}2Q\rho)}$.
As stated before, these states do not
contribute to the outgoing flux and their importance in the
construction of the scattering state diminishes very rapidly for large values 
of $\rho$. The approximation introduced for $\rho>\rho_0$
in the application of the boundary condition to the states with 
$\mu>\overline{N}_m$ has been checked by increasing the value
of the matching radius. In the cases considered here
the solutions obtained for the $S$-matrix show a complete stability
for values of the matching radius $\rho_0>100$ fm.

Let us define $|\mu,m>$ to be a correlated totally antisymmetric element
of the expansion basis. Here $\mu$ indicates the correlated HH state
${\cal Q}_\mu(\rho,\Omega)$ and $m=1,....,M$ indicates the Laguerre
polynomial $L^{(5)}_m(z)$ or, for $m=M+1$, the auxiliary function
$\overline\omega_\mu$. In terms of these basis elements
the internal part of the wave function is

\be
   \label{eq:exp1}
\Psi_C=\sum_{\mu,m}A_\mu^m|\mu,m> \ .
\ee
The variation of the functional $[\;{}^JS^{SS}_{LL}]$ with respect
to the linear parameters leads to the following set of linear equations
\be
   \label{eq:exp2}
  \sum_{\mu',m'}A^{m'}_{\mu'}<\mu,m|H-E|\mu',m'>=D^\lambda_{\mu,m} \ ,
\ee
where the inhomogeneous term is

\be
   \label{eq:exp3}
  D^\lambda_{\mu,m}= \sum_j<\mu,m|H-E|\Omega^\lambda_{LSJ}(\x_i,\y_i)> \ .
\ee

The first order solution of the $S$-matrix is obtained solving the 
algebraic equations

\be
   \label{eq:exp4}
\sum_{L''S''} \;^JS^{SS'}_{LL'}\; X^{S'S''}_{L'L''}=Y^{SS'}_{LL'} \ ,
\ee
with the coefficients X and Y defined to be

\bea
   \label{eq:exp5}
  X^{SS'}_{LL'}&= <\Omega^1_{LSJ}+\Psi^1_{LSJ}|H-E|\Omega^1_{LSJ}>  
  \nonumber \\
  Y^{SS'}_{LL'}&= <\Omega^0_{LSJ}+\Psi^0_{LSJ}|H-E|\Omega^0_{LSJ}> \ , 
\eea
where $\Psi^\lambda_{LSJ}$ is constructed using the solution of 
eq.(\ref{eq:exp2}) with
the corresponding inhomogeneous term. The second order estimate
$[\;^JS^{SS'}_{LL'}]$ is obtained replacing the first order solution in
eq.(\ref{eq:ckohn}).

Finally, due to flux conservation the following
condition has to be satisfied between the matrix elements of the
elastic $S$-matrix and the coefficients of the outgoing breakup waves:
\begin{equation}
\sum_{S'L'}|{}^JS^{SS'}_{LL'}|^2+\sum_\mu |b_\mu|^2 =1 \ .
\end{equation} 
The coefficients $b_\mu$, which are defined in eq.(\ref{eq:asy2}),
are the linear parameters $A_\mu^{M+1}$ of eq.(\ref{eq:M}).
The above relation
allows the calculation of the total breakup cross section 
from the elastic $S$-matrix elements, as has been recently discussed in 
ref.~\cite{KBV00}.

\section{Numerical Results}

In order to study the solution of eq.(\ref{eq:siste}) by means of the
expansion given in eq.(\ref{eq:M}), we have first calculated the phase-shift 
and inelasticity parameters for n-d and p-d scattering using the spin-dependent 
s-wave potential of Malfliet and Tjon. The results are presented in Table I 
for two energy values, $E_{lab}=14.1$ and $42.0$ MeV. The calculations have
been done using $N_\alpha=8$ hyperspherical polynomials per channel,
as in the case already studied in ref.~\cite{KVR97} where
the set of equations SE1 was solved using the finite difference technique.
Moreover, since the potential is central,
the phase-shifts $^{2S+1}\delta_L$ and inelasticities $^{2S+1}\eta_L$
do not depend on the total angular momentum $J$. Only the case $L=0$
has been considered and the results
are given in Table I for increasing values of the number of 
Laguerre polynomials $M$. For the sake of comparison the results of 
ref.~\cite{KVR97} 
are reported as well as the benchmark results of ref.\cite{bench2}
obtained by solving the Faddeev equations in
configuration space (Los Alamos group) and momentum space (Bochum group).
We observe a very fast convergence with
$M$ and, in general, $16$ to $20$ polynomials are enough to obtain
the phase-shift and mixing parameters with four digit accuracy. With the
number of Laguerre polynomials that has been taken into account a very low
dependence on the nonlinear parameter $\gamma$ has been observed. 
In fact the results
reported here do not change for variations of the parameter in the
range $1.5$  fm$^{-1}\leq\gamma\leq 2.5$ fm$^{-1}$. Moreover, the
dimension of the matrices involved in the solution is one order of 
magnitude smaller than that used in~\cite{KVR97}.

The case of realistic interactions has been considered in
refs.~\cite{KRV99,VKR00} where the AV18 interaction has been used to 
calculate p-d scattering at $E_{lab}=5$ and $10$ MeV. In particular,
in ref.~\cite{VKR00} the convergence of the phase-shift and mixing 
parameters for the state $J=1/2^+$ has been studied by increasing the number 
of angular-spin-isospin channels. The convention discussed in ref.~\cite{huber}
has been adopted in the parametrization of the $S$--matrix in terms 
of phase-shift and mixing parameters.
In order to illustrate the variation of these parameters with
energy the doublet and quartet $S$, $P$ and $D$ phases, denoted as 
$^{2S+1}L_J$, are reported in Fig.1 as well as the mixing parameters
$\eta_{1/2+}$, $\eta_{3/2+}$, $\epsilon_{1/2-}$ and $\epsilon_{3/2-}$. 
Both the real and imaginary parts are shown.
It is interesting to notice that the splitting in the real part
of the phases with equal spin $S$ and angular momentum $L$ but different 
$J$, increases with energy. Conversely, the imaginary parts of the phases,
which are related to the inelasticity of a state with a given value of
$J$, reveal a tiny splitting. After summing all the contributions, 
the total breakup cross section can be obtained, as is discussed
in the next section.

\section{p-d cross sections}

The calculation of scattering observables using the present variational
technique is based on the estimate of the elastic 
$S$-matrix for all states with $J \le J_{Max}$.
Each observable is obtained from a trace operation after the
evaluation of the transition matrix, following the formalism of
Seyler~\cite{seyler}. The value of $J_{Max}$ has been chosen by requiring
that partial waves with $J>J_{Max}$ give negligible
contributions to all the observables considered. In the present work
results for cross sections, vector and tensor analyzing powers up to
$E_{lab}=28$ MeV are presented, and correspondingly
the value $J_{Max}=19/2$ has been found to be appropriate.

Let us start with the analysis of the p-d cross sections.
For p-d scattering the total breakup cross section 
accounts for all possible configurations in which all three particles
are moving away from each other. 
Its expression can be given in terms of the elastic 
$S$-matrix~\cite{KBV00},

\begin{equation}
\sigma_b(p-d)={\pi \over k^2}{1 \over 6}
     \sum_J(2J+1) tr\{I_J-S_JS_J^\dagger\} ,
\label{eq:sigmab}
\end{equation} 
where $k^2=2\mu E_{cm}/\hbar^2$ ($\mu$ is the nucleon-deuteron
reduced mass) and
$I_J$ is the $3 \times 3$ identity matrix, except for $J=1/2$
which is the $2 \times 2$ identity matrix. The quantity $S_J$ is the elastic
$S$-matrix for the state $J$. The sum runs over all possible values
of $J$ and parity (the sum over the two parities is implied).
In principle the sum runs from $J=0$ to infinity, but there is a rapid
convergence since each $S_J$ matrix becomes closer to unitary as $J$ 
increases. In Fig.2 the theoretical prediction for $\sigma_b(p-d)$
is given together with the two sets of data available in the literature.
The first data set corresponds to energies just above the DBT~\cite{GM59} 
whereas the second starts at 20 MeV~\cite{Carl73}.
The solid line is the AV18 prediction and
is found to be in reasonable agreement with both sets of data.
The inclusion of the URIX potential does not produce appreciable
modifications and both results, with and without the
inclusion of the 3NF, nearly coincide.
The low sensitivity to the 3NF can be understood by noticing that the
contribution to $\sigma_b$ comes from a balance between the
spin factor $2J+1$ and the quantity
$tr\{I_J-S_jS^\dagger_J \}$ which can be considered as a measurement
of the inelasticity of the state (divided by $tr\{I_J\}$). Above
$5$ MeV the state $J=3/2^-$ gives by far the main contribution to the
observable~\cite{KBV00}. The state $J=1/2^+$, which is appreciably modified by 
the 3NF, has the largest inelasticity, but due to a small spin factor 
it gives a contribution of the same order as other states that are much less 
``inelastic'' and modified slightly by the 3NF. The final result after summing
up all these contributions is that the small (but sizeable) 
effect of the 3NF on $J=1/2+$ has no impact in $\sigma_b$.

Regarding the elastic p-d differential cross section, a huge amount of high 
quality data has been collected during the past years. Low energy 
measurements have been taken recently at TUNL at different energy values
below $E_{lab}=1$ MeV~\cite{brune,black,wood}. An analysis of the quality 
in the description of these data has been performed using the AV18 and 
the AV18+URIX interactions~\cite{kw00,carl01}. It was shown that 3NF 
effects can be revealed through a $\chi^2$ analysis of the data. 
Essentially these effects are related to a correct description of
the $^3$He binding energy. In fact, using the AV18+URIX interaction it is
possible to describe the p-d differential cross section at
$E_{lab}=1$, $2$ and $3$ MeV with a $\chi^2$ per datum ($\chi^2_N$)
close to one. This value increases significantly when the
AV18 potential is considered alone. 
The agreement between the theoretical and experimental differential cross 
section worsens, though not dramatically, as the energy increases.
For example, at $E_{lab}=135$ MeV a value of $\chi^2_N=16.9$ $(225.2)$ 
was recently obtained with (without) the
inclusion of a 3NF~\cite{sakai}. Again the inclusion of a 3NF reduces
the $\chi^2$ per datum considerably. 

The results obtained for the p-d differential cross section
are given in Fig.3 for nine values of the energy,
$E_{lab}=5,7,9,10,12,14,16,18,22.7,28$ MeV.
For each energy three curves are shown corresponding to 
calculations using the AV18 potential (solid line), the AV18+URIX
potential (dotted line), and calculations for n-d scattering
using the AV18 potential (dashed line). The theoretical predictions
are compared to the experimental data of 
refs.~\cite{sagara,bruebler83,hatanaka84}, with the exception of the
calculations at $16$ MeV which are compared to data obtained at a slightly
different energy ($16.5$ MeV).
The analysis of the results at the different energies shows
that 3NF effects are small in this 
energy range and the AV18 and AV18+URIX curves practically overlap
each other. A more quantitative analysis at $E_{lab}=18$ MeV
gives $\chi^2_N=11$ using the AV18 interaction and nearly the same value
for AV18+URIX. Coulomb effects are mainly observed at forward and backward
angles whilst they are strongly reduced at the minimum. 
Tiny Coulomb effects at the minimum are confirmed by comparing
n-d to p-d data, as can be seen in ref.~\cite{report}.

As far as the agreement between theory and experiment is concerned,
the situation for the differential cross section above the DBT is
different when compared to what has been observed below the DBT. 
As mentioned before, at very low energies the differential
cross section can be described with $\chi^2_N \approx 1$ using AV18+URIX,
while when the AV18 is used alone a substantially worse result, 
$\chi^2_N > 10$, is obtained. In fact, the AV18 curve
remains above the data points all over the angular distribution.
As the energy increases the tendency for the AV18 curve is to go
below the data at the minimum.
This problem is appreciable already at $28$ MeV, as can be seen in the
last panel of Fig.3. Around 20 MeV
the AV18+URIX curve starts to rise above the AV18 curve and
closer to the data. This effect is clearly shown in ref.~\cite{witala}
where Faddeev calculations using several NN and 3NF interactions have 
been compared to the data at $3$, $65$, $135$ and $190$ MeV.
In the energy range analyzed here we observe that there is one energy, 
around 18 MeV, where the AV18 and AV18+URIX curves mostly overlap.
In order to analyze further this behavior, in Table II, the values at
the minimum of the p-d cross section calculated with AV18 and AV18+URIX
are compared to the data. The corresponding values of the AV18 n-d cross 
section are also given in order to have a quantitative idea of the size of
the Coulomb effects.

\section{Polarization observables}

The vector and tensor analyzing powers are examples of polarization
observables. There is a large amount of p-d and d-p data for the 
vector analyzing powers $A_y$ and $iT_{11}$ as well as for the
tensor analyzing powers $T_{20},T_{21},T_{22}$. The study of these
observables is important because they are sensible to the non-central terms of 
the nuclear interaction. These terms are responsible for small 
components in the wave function which in general are less known. Therefore, 
the accuracy shown by the modern interactions when reproducing the
vector and tensor analyzing powers in the three-nucleon system 
gives important information about parts of the nuclear interaction not
completely under control. As is well known, in the low energy region
the vector analyzing powers
are heavily underpredicted by all modern NN interactions and the origin 
of this discrepancy is not yet completely understood. Possible ways for
solving this puzzle have recently been investigated, based on the
inclusion of new terms in the three-nucleon potential~\cite{kie99,canton00}
or on a new NN potential obtained from chiral perturbation 
theory~\cite{epelbaum}. These studies represent only a first step in 
the understanding of the puzzle and further investigations and
refinements of the models are needed.
A similar underprediction of the proton analyzing power $A_y$ has been
found in calculations on p-$^3$He scattering, as was recently
pointed out~\cite{vkrkg00}. Therefore, a solution to the puzzle
should concern both the 3N and 4N systems.

In the present paper, we will discuss the quality of the description of 
the vector and tensor polarization observables achieved by the AV18 and 
the AV18+URIX interactions in p-d scattering up to $28$ MeV. In Fig.4
the results for $A_y$ are given for the same nine energy values given
in Fig.3. The three curves correspond to the p-d $A_y$ calculated
using AV18 (solid line) and AV18+URIX (dotted line), and the n-d $A_y$
calculated using AV18 (dashed line). 
The calculations are compared to
data from ref.~\cite{sagara} at $E_{lab}=5,7,9,10,12,16,18$ MeV,
and from ref.~\cite{bruebler83} at $E_{lab}=22.7$ MeV.
As expected, Coulomb effects
are appreciable in all the energy range. Below $18$ MeV the effects are
appreciable at the maximum. Above $18$ MeV the shape of $A_y$ changes and
a clear minimum appears, where Coulomb effects can be observed. 
The order of magnitude of these effects is $15$\%.
Instead, 3NF effects are not so important. This is a characteristic
of the Urbana potential that modifies the quartet $P$-waves (which produces
the main contribution to $A_y$ below $30$ MeV) in such a way that there is 
a cancellation among the different contributions, so
the global effect on the observable is small. A similar
analysis holds for $iT_{11}$, shown in Fig.5,
since these two observables have rather similar structures. 
The calculations for $iT_{11}$ are compared to data from ref.~\cite{sagara} 
at $E_{lab}=5,7,9$ MeV, from ref.~\cite{bruebler83} at 
$E_{lab}=10,12,16.5,22.7$ MeV and from ref.~\cite{hatanaka84} at 
$E_{lab}=28$ MeV, taking care again that at $16$ MeV the comparison is to 
data obtained at a slightly different energy ($16.5$ MeV).
As a difference between the proton and deuteron analyzing powers,
we observed that Coulomb and 3NF effects are of the same size at the
minimum of $iT_{11}$ above $16$ MeV.

In Figs.6-8 the tensor observables $T_{20},T_{21},T_{22}$ are given, 
respectively. 
As before, the three curves correspond to the p-d $T_{ij}$ calculated
using AV18 (solid line) and AV18+URIX (dotted line), and the n-d $T_{ij}$
calculated using AV18 (dashed line). 
The calculations are compared to data from ref.~\cite{sagara} 
at $E_{lab}=5,7,9$ MeV, from ref.~\cite{bruebler83} at 
$E_{lab}=10,12,16.5,22.7$ MeV and from ref.~\cite{hatanaka84} at 
$E_{lab}=28$ MeV.
As a general trend, the agreement with the data for the tensor
observables is better than for the vector observables. Coulomb effects
are appreciable in the three observables at low energies. As the
energy increases the inclusion of the Coulomb interaction in the analysis
of the tensor observables is less important, mostly for $T_{20}$
and $T_{22}$. The case of $T_{21}$ is of particular interest since
Coulomb effects are still appreciable at $28$ MeV.
This observation suggests that comparisons of d-p data to
calculations where the Coulomb interaction has been neglected should be
done with caution.
The effect of the 3NF is somehow contradictory since
in some cases its inclusion improves the description of the observables
but in other cases it does not. For example,
a net improvement is obtained in the description of the minimum of 
$T_{22}$ below 12 MeV. Also the maximum of $T_{20}$ and $T_{21}$ is better
described with the AV18+UR potential. Conversely the description of
the minimum of $T_{20}$ and $T_{21}$ is better described by the
AV18 potential alone. The case of $T_{21}$ is again of
interest since 3NF effects seem to be bigger in this tensor
observable than in the others.
Unfortunately experimental data for $T_{21}$ are not available at
all the energies.

In order to give a quantitative estimation of the agreement between the
theoretical calculations and the measurements, the $\chi^2_N$
for the polarization observables is presented in Table III.  
In the first row of the table, the
$\chi^2_N$ is given with respect to a recent measurement performed at 
1 MeV~\cite{wood} and in the second row with respect to the measurements
of ref.~\cite{shimizu} at $E_{lab}=3$ MeV. These two energies are below
the DBT and are useful for analyzing the trend of $\chi^2$ starting at
low energies. By inspection of the table, the manifestation of the
$A_y$ puzzle is evident since the $\chi^2_N$ for the vector observables 
is a few hundreds at low energy. Above $18$ MeV $A_y$ and $iT_{11}$
change shape being closer to the shape of $T_{21}$ with a pronounced
minimum followed by a maximum. After that energy the values of
$\chi^2_N$ decrease in such a way that, at the last energy, 
vector and tensor observables have similar values which are of the order of
one tenth. These final values are
comparable to those ones obtained recently in ref.~\cite{sakai}
at 135 MeV.

\section{Conclusions}

In the present paper we have studied p-d elastic
scattering above the DBT, up to $E_{lab}=28$ MeV.
The differential cross section and the total breakup cross section
as well as the vector and tensor analyzing powers have been calculated using
one of the modern NN interactions, namely the AV18 potential. In order to
evaluate 3NF effects the three-nucleon potential of Urbana has been 
taken into account. The effects of the Coulomb interaction
has been considered in the framework of the complex Kohn Variational
Principle. 

The internal part of the p-d scattering wave function has been 
expanded in terms of the PHH basis. The KVP
has been applied to obtain a set of differential equations
for the hyperradial functions. The set has been solved imposing
outgoing boundary conditions at a certain value of the hyperradius
$\rho=\rho_0$ and expanding the hyperradial functions in the
region $[0,\rho_0]$ in Laguerre polynomials plus an 
auxiliary oscillating function.
The solution should not depend on the value of $\rho_0$ providing
that for $\rho>\rho_0$ the asymptotic behavior has been reached.
Such a technique has proved to be adequate since the results from 
ref.~\cite{KVR97}
has been reproduced as well as the benchmark of ref.~\cite{bench2}. 

The calculations have been extended to all states and parities with
$J\le 19/2$, corresponding to nine energies up to $E_{lab}=28$ MeV. 
The elastic $S$-matrix has then been used to calculate the observables 
of interest and compare them to the data. Moreover, the corresponding
observables for n-d scattering, where the Coulomb interaction is absent,
have been calculated too. From the analysis of the results
some conclusions can be drawn about the capability of
the AV18 and AV18+URIX interactions to reproduce the data. 
A quantitative measure of the agreement achieved by the theory in the
description of the data has been given through a $\chi^2$ analysis.
No appreciable 3NF effects have been observed in the total breakup
cross section and small effects appear in the differential cross section. 
However, results using the AV18+URIX model produce a lower
$\chi^2_N$ value than those where the AV18 interaction has been used alone. 
At low energies this is a
manifestation of the correct description of the $^3$He binding energy 
by the AV18+URIX interaction. But as the energy increases, there
is a different sensitivity to the 3NF. In particular, at the highest energy
considered, $E_{lab}=28$ MeV, the AV18+URIX differential cross section
is above the AV18 differential cross section, reversing the order
observed at lower energies.  This situation becomes much more evident 
for example at $E_{lab}=60$ MeV~\cite{witala}. In order to further analyze
this behavior in Table II the minimum of the AV18 and AV18+URIX
cross sections are compared to the data. The minimum of
the n-d differential cross section calculated with the AV18 potential
is also given in order to estimate Coulomb effects. The results
of the table are useful for analyzing what has been called the 
``Sagara discrepancy''~\cite{sagarad}.

In the case of vector and tensor analyzing powers the Urbana 3NF has little
impact below $30$ MeV. The centrifugal barrier
is still strong and there is not too much sensitivity to the short
range part of the interaction. 
Conversely there are important Coulomb effects.
In order to improve the description of
the vector analyzing powers new terms could be considered in the
three-nucleon potential. In such a case both the vector and
the tensor analyzing powers should improve as well.

The present picture of the 3N scattering from the theoretical
point of view is the following. At low
energies there is a large underprediction of the vector analyzing powers
whereas the differential cross section and 
tensor analyzing powers are well described. Up to $30$ MeV we see an 
improvement in $A_y$ and $iT_{11}$ indicating that the $A_y$ puzzle is a 
low energy problem. On the other hand,
a progressive deterioration in the description of the cross section
and tensor observables is revealed through a $\chi^2$ analysis.
For energies above $30$ MeV we can refer to the very recent work of 
ref.~\cite{witala} and we see that this picture remains essentially the
same up to very high energies (around $135$ MeV). Above this energy
a number of new conflicts appears.
 
At present a few realistic local and non-local NN interactions have
been determined by accurately fitting the two nucleon scattering
observables. All of them give rise to the $A_y$ puzzle in the
three-nucleon system. It seems difficult to derive a new NN interaction, 
which still accurately fits the two-nucleon scattering, and correctly 
describes the N-d $A_y$ at low energies~\cite{friar}. So,
if this is the case, a possible solution to the puzzle should come from
an improvement of the currently used 3NF models. Indeed, in accordance with
chiral perturbation theory, these 3NF's include those terms of larger 
magnitude. On the other hand,
the $A_y$ puzzle can be solved by rather small changes in certain
P-wave phase-shifts, which can be obtained by adding a small term to the
three-nucleon potential~\cite{kie99}. However, as the energy increases other
discrepancies arise  in the polarization observables that could
have different origins. Hence, since the three-nucleon continuum
can be calculated at present with great accuracy, it is reasonable to
expect that the actual 3NF models can be adjusted to describe the
3N data. The necessary calculations will require
large computing time, but is our opinion that this project 
will certainly help to understand the long unsolved problems
in low energy nuclear physics.

\begin{acknowledgements}
One of us (A.K.) would like to thank W. Tornow, E. Ludwig, H. Karwowski and 
C. Brune for useful discussions, and the Triangle Universities
Nuclear Laboratory for hospitality and support where part of this 
work was performed. The authors wish to thank L. Lovitch for
a critical reading of the manuscript.
\end{acknowledgements}

\newpage
\begin{table}
\caption{Phase--shift and mixing parameters for different values of the 
number $M$ of the Laguerre
polynomials used in the expansion of the hyperradial functions. The
$s$--wave potential of Malfliet--Tjon has been considered.}
\label{tab:phasec}
\begin{tabular}{c|cc|cc}
\hline
 & \multicolumn{4}{c}{n-d at $ E_n=14.1$ MeV} \\
\hline
 $M$       & $^2\delta_0$ & $^2\eta_0$ & $^4\delta_0$ & $^4\eta_0$ \\ 
\hline
  4        & 104.44       & 0.4672     &  68.993      &  0.9669    \\
  8        & 105.33       & 0.4663     &  68.963      &  0.9774    \\
  12       & 105.42       & 0.4658     &  68.951      &  0.9781    \\
  16       & 105.49       & 0.4646     &  68.952      &  0.9782    \\
  20       & 105.48       & 0.4648     &  68.952      &  0.9782    \\
  24       & 105.48       & 0.4649     &  68.952      &  0.9782    \\
  28       & 105.48       & 0.4649     &  68.952      &  0.9782    \\
\hline
ref.[27]   & 105.50       & 0.4649     &  68.95       &  0.9782    \\
Los Alamos & 105.48       & 0.4648     &  68.95       &  0.9782    \\
Bochum     & 105.50       & 0.4649     &  68.96       &  0.9782    \\
\hline
\hline
 & \multicolumn{4}{c}{p-d at $ E_p=14.1$ MeV} \\
\hline
  4        & 107.37       & 0.5006     &  71.665      &  0.9654    \\
  8        & 108.34       & 0.4984     &  72.615      &  0.9799    \\
  12       & 108.42       & 0.4988     &  72.602      &  0.9794    \\
  16       & 108.45       & 0.4984     &  72.602      &  0.9795    \\
  20       & 108.43       & 0.4984     &  72.604      &  0.9795    \\
  24       & 108.44       & 0.4984     &  72.604      &  0.9795    \\
  28       & 108.44       & 0.4984     &  72.604      &  0.9795    \\
\hline
 ref.[27]  & 108.43       & 0.4984     &  72.60       &  0.9795    \\
\hline
 & \multicolumn{4}{c}{n-d at $ E_n=42.0$ MeV} \\
\hline
 $M$       & $^2\delta_0$ & $^2\eta_0$ & $^4\delta_0$ & $^4\eta_0$ \\ 
\hline
  4        &  42.198      & 0.4575     &  38.218      &  0.8917    \\
  8        &  41.818      & 0.4934     &  37.680      &  0.9028    \\
  12       &  41.147      & 0.5009     &  37.607      &  0.9016    \\
  16       &  41.271      & 0.5010     &  37.724      &  0.9027    \\
  20       &  41.332      & 0.5020     &  37.723      &  0.9031    \\
  24       &  41.340      & 0.5022     &  37.722      &  0.9033    \\
  28       &  41.341      & 0.5022     &  37.722      &  0.9033    \\
\hline
ref.[27]   &  41.33       & 0.5026     &  37.71       &  0.9034    \\
Los Alamos &  41.34       & 0.5024     &  37.71       &  0.9035    \\
Bochum     &  41.37       & 0.5022     &  37.71       &  0.9033    \\
\hline
 & \multicolumn{4}{c}{p-d at $ E_n=42.0$ MeV} \\
\hline
 $M$       & $^2\delta_0$ & $^2\eta_0$ & $^4\delta_0$ & $^4\eta_0$ \\ 
\hline
  4        &  38.048      & 0.3988     &  39.937      &  0.8738    \\
  8        &  44.701      & 0.4961     &  40.554      &  0.9103    \\
  12       &  43.479      & 0.5079     &  39.844      &  0.9011    \\
  16       &  43.618      & 0.5059     &  39.934      &  0.9036    \\
  20       &  43.660      & 0.5055     &  39.947      &  0.9043    \\
  24       &  43.667      & 0.5056     &  39.947      &  0.9046    \\
  28       &  43.667      & 0.5056     &  39.947      &  0.9046    \\
\hline
ref.[27]   &  43.65       & 0.5058     &  39.94       &  0.9047    \\
\end{tabular}
\end{table}

\newpage
\begin{table}[h]
\caption{The minimum of the n-d and p-d differential cross sections (in mb/sr)
 at different energies,
calculated using the AV18 and AV18+UR potential models. Experimental
data for the p-d cross section are from 
refs.~\protect\cite{hatanaka84,sagara,bruebler83,kw00}}
\label{table:2}
\begin{tabular}{l|cccc}
\hline
Energy   & AV18(n-d)& AV18(p-d)& AV18+UR(p-d)& exp. \\
\hline
1 MeV    &   148.9  & 177.5   &  170.7   &  170.2$\pm$1.3  \\
3 MeV    &   92.7   &  96.0   &   92.3   &   91.1$\pm$0.7  \\
5 MeV    &   53.1   &  56.2   &   53.8   &   52.7$\pm$0.4  \\
7 MeV    &   32.9   &  35.3   &   33.9   &   32.9$\pm$0.2  \\
9 MeV    &   21.3   &  23.1   &   22.1   &   21.8$\pm$0.2  \\
10 MeV   &   17.3   &  18.9   &   18.2   &   18.0$\pm$0.2  \\
12 MeV   &   11.8   &  12.8   &   12.5   &   12.2$\pm$0.1  \\
16 MeV   &   6.0    &   6.5   &   6.4    &    6.2$\pm$0.1  \\
18 MeV   &   4.5    &   4.9   &   4.9    &    4.7$\pm$0.1  \\
22.7 MeV &   2.7    &   2.8   &   2.9    &    2.89$\pm$0.03  \\
28 MeV   &   1.8    &   1.8   &   1.9    &    2.19$\pm$0.02  \\
\hline
\end{tabular}
\end{table}
\newpage

\begin{table}[h]
\caption{$\chi^2$ per datum obtained in the description of the vector and
tensor analyzing powers at several energies using the AV18 and AV18+UR 
potentials}
\label{table:3}
\begin{tabular}{ll|ccccc}
\hline
Energy   & potential & $A_y$&$iT_{11}$&$T_{20}$&$T_{21}$&$T_{22}$ \\
\hline
1 MeV    & AV18      & 276  & 112     & 3.5    & 4.5    &  2.8    \\
         & AV18+UR   & 190  & 61      & 1.0    & 2.5    &  0.7    \\
\hline
3 MeV    & AV18      & 313  & 205     & 4.8    & 6.7    &  12     \\
         & AV18+UR   & 271  & 144     & 5.4    & 11     &  2.4    \\
\hline
5 MeV    & AV18      & 211  & 99      & 6.8    & 12     &  7.8    \\
         & AV18+UR   & 186  & 59      & 26     & 36     &  1.5    \\
\hline
7 MeV    & AV18      & 303  &  90     & 19     & 38     &  1.9    \\
         & AV18+UR   & 239  &  56     & 40     & 81     &  4.2    \\
\hline
9 MeV    & AV18      & 292  & 165     & 42     & 70     &  38     \\
         & AV18+UR   & 218  & 134     & 63     & 86     &  7.2    \\
\hline
10 MeV   & AV18      & 288  & 29      & 10     & 6.2    &  24     \\
         & AV18+UR   & 224  & 23      & 13     & 6.1    &  7.6    \\
\hline
12 MeV   & AV18      & 313  & 50      & 19     &  --    &  39     \\
         & AV18+UR   & 227  & 34      & 16     &  --    &  22     \\
\hline
16 MeV   & AV18      & 296  & 80      & 114    &  --    &  70     \\
         & AV18+UR   & 246  & 61      & 139    &  --    &  48     \\
\hline
18 MeV   & AV18      & 293  &  --    &   --    &  --    &  --     \\
         & AV18+UR   & 250  &  --    &   --    &  --    &  --     \\
\hline
22.7 MeV & AV18      &  78  &  89     & 44     &   --   &  24     \\
         & AV18+UR   &  72  &  61     & 59     &   --   &  17     \\
\hline
28 MeV   & AV18      &  --  & 19      & 10     & 7.1    &  11     \\
         & AV18+UR   &  --  & 13      & 10     & 11     &  8.5    \\
\hline
\end{tabular}
\end{table}

\newpage

Figure Captions

Fig.1. Phase--shift and mixing parameters (in degrees) as a function
of energy. (a) $^2S_{1/2}$ and $^4S_{3/2}$, their real and imaginary
parts are indicated by $(\bigcirc,+)$ and $(\Box,\times)$, respectively.
(b) $^2P_{1/2}$ and $^2P_{3/2}$, their real and imaginary
parts are indicated by $(\bigcirc,+)$ and $(\Box,\times)$, respectively.
(c) $^2D_{3/2}$ and $^2D_{5/2}$, their real and imaginary
parts are indicated by $(\bigcirc,+)$ and $(\Box,\times)$, respectively.
(d) $^4P_{1/2}$, $^4P_{3/2}$ and $^4P_{5/2}$, their real and imaginary
parts are indicated by $(\bigcirc,+)$, $(\Box,\times)$
and $(\Diamond,*)$, respectively.
(e) $^4D_{1/2}$, $^4D_{3/2}$, $^4D_{5/2}$ and $^4D_{7/2}$, their real parts
are indicated by $\bigcirc$, $\Box$, $\Diamond$ and $\triangle$, respectively.
Only the imaginary part of $^4D_{1/2}$ is given (+). 
(f) Mixing parameters $\eta_{1/2+}$, $\eta_{3/2+}$, $\epsilon_{1/2-}$ and
$\epsilon_{3/2-}$, their real parts are indicated by
$\bigcirc$, $\Box$, $\triangle$ and $\bigtriangledown$, respectively.
The imaginary parts of $\epsilon_{1/2-}$ and $\epsilon_{3/2-}$
are indicated by + and $\times$, respectively. The imaginary
parts of $\eta_{1/2+}$ and $\eta_{3/2+}$ are close to zero and are not
shown.

Fig.2. The p-d total breakup cross section $\sigma_b$ below 30 MeV
calculated using the AV18 interaction. The
experimental results of Gibbons and Macklin \cite{GM59} (open triangles)
and Carlson {\sl et al.} \cite{Carl73} (open circles) are given for the
sake of comparison.

Fig.3. N-d differential cross section up to $28$ MeV. Calculations
are shown for p-d scattering
using the AV18 (solid line) and AV18+UR (dotted line) potentials 
and for n-d scattering using the AV18 potential (dashed line).
Data are from ref.~\cite{sagara} at $E_{lab}=5,7,9,10,12,16,18$ MeV,
from ref.~\cite{bruebler83} at $E_{lab}=22.7$ MeV and from
ref.~\cite{hatanaka84} at $E_{lab}=28$ MeV.

Fig.4. Nucleon vector analyzing power $A_y$ up to $28$ MeV. Calculations
are shown for p-d scattering
using the AV18 (solid line) and AV18+UR (dotted line) potentials 
and for n-d scattering using the AV18 potential (dashed line).
Data are from ref.~\cite{sagara} at $E_{lab}=5,7,9,10,12,16,18$ MeV,
and from ref.~\cite{bruebler83} at $E_{lab}=22.7$ MeV.

Fig.5. Deuteron vector analyzing power $iT_{11}$ up to $28$ MeV.
Calculations are shown for p-d scattering
using the AV18 (solid line) and AV18+UR (dotted line) potentials 
and for n-d scattering using the AV18 potential (dashed line).
Data are from ref.~\cite{sagara} at $E_{lab}=5,7,9$ MeV,
from ref.~\cite{bruebler83} at $E_{lab}=10,12,16.5,22.7$ MeV and from
ref.~\cite{hatanaka84} at $E_{lab}=28$ MeV.

Fig.6. Tensor analyzing power $T_{20}$ up to $28$ MeV.
Calculations are shown for p-d scattering
using the AV18 (solid line) and AV18+UR (dotted line) potentials 
and for n-d scattering using the AV18 potential (dashed line).
Data are from ref.~\cite{sagara} at $E_{lab}=5,7,9$ MeV,
from ref.~\cite{bruebler83} at $E_{lab}=10,12,16.5,22.7$ MeV and from
ref.~\cite{hatanaka84} at $E_{lab}=28$ MeV.

Fig.7. Tensor analyzing power $T_{21}$ up to $28$ MeV.
Calculations are shown for p-d scattering
using the AV18 (solid line) and AV18+UR (dotted line) potentials 
and for n-d scattering using the AV18 potential (dashed line).
Data are from ref.~\cite{sagara} at $E_{lab}=5,7,9$ MeV,
from ref.~\cite{bruebler83} at $E_{lab}=10$ MeV and from
ref.~\cite{hatanaka84} at $E_{lab}=28$ MeV.

Fig.8. Tensor analyzing power $T_{22}$ up to $28$ MeV.
Calculations are shown for p-d scattering
using the AV18 (solid line) and AV18+UR (dotted line) potentials 
and for n-d scattering using the AV18 potential (dashed line).
Data are from ref.~\cite{sagara} at $E_{lab}=5,7,9$ MeV,
from ref.~\cite{bruebler83} at $E_{lab}=10,12,16.5,22.7$ MeV and from
ref.~\cite{hatanaka84} at $E_{lab}=28$ MeV.

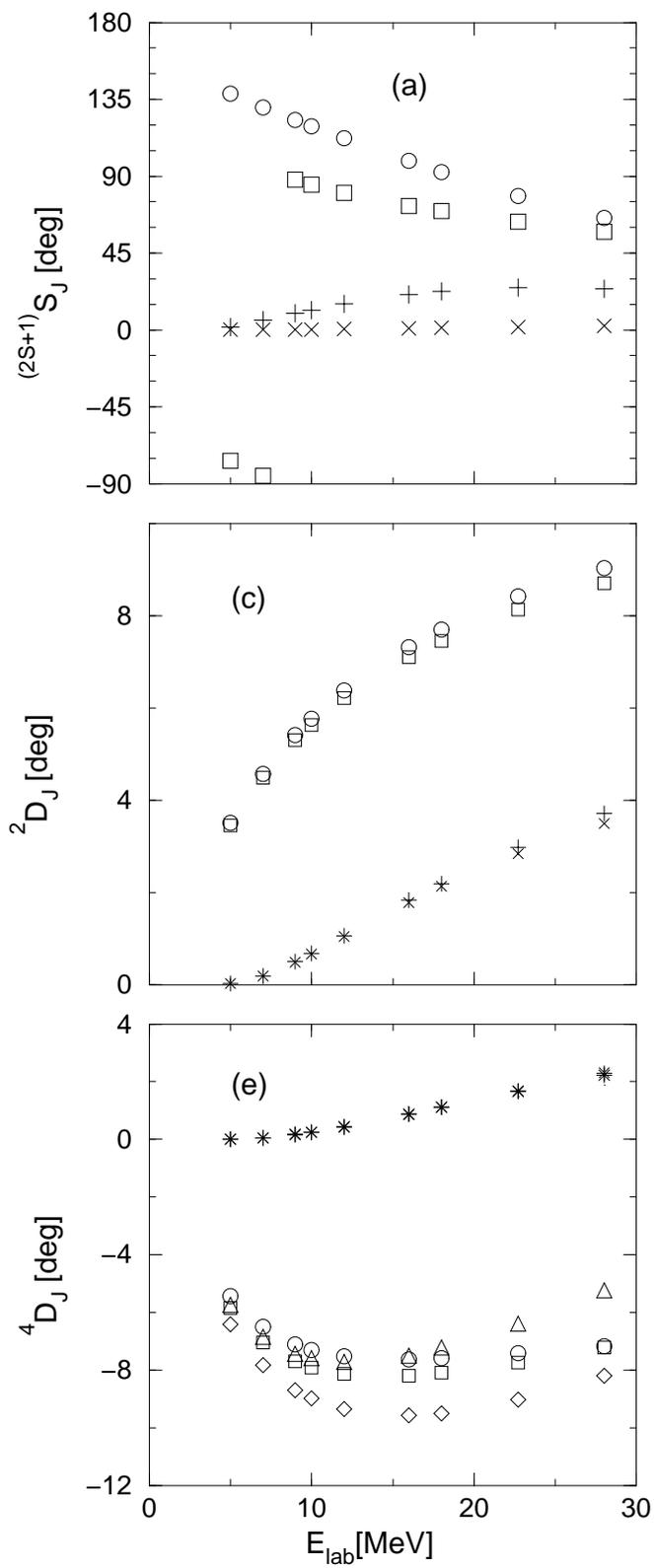
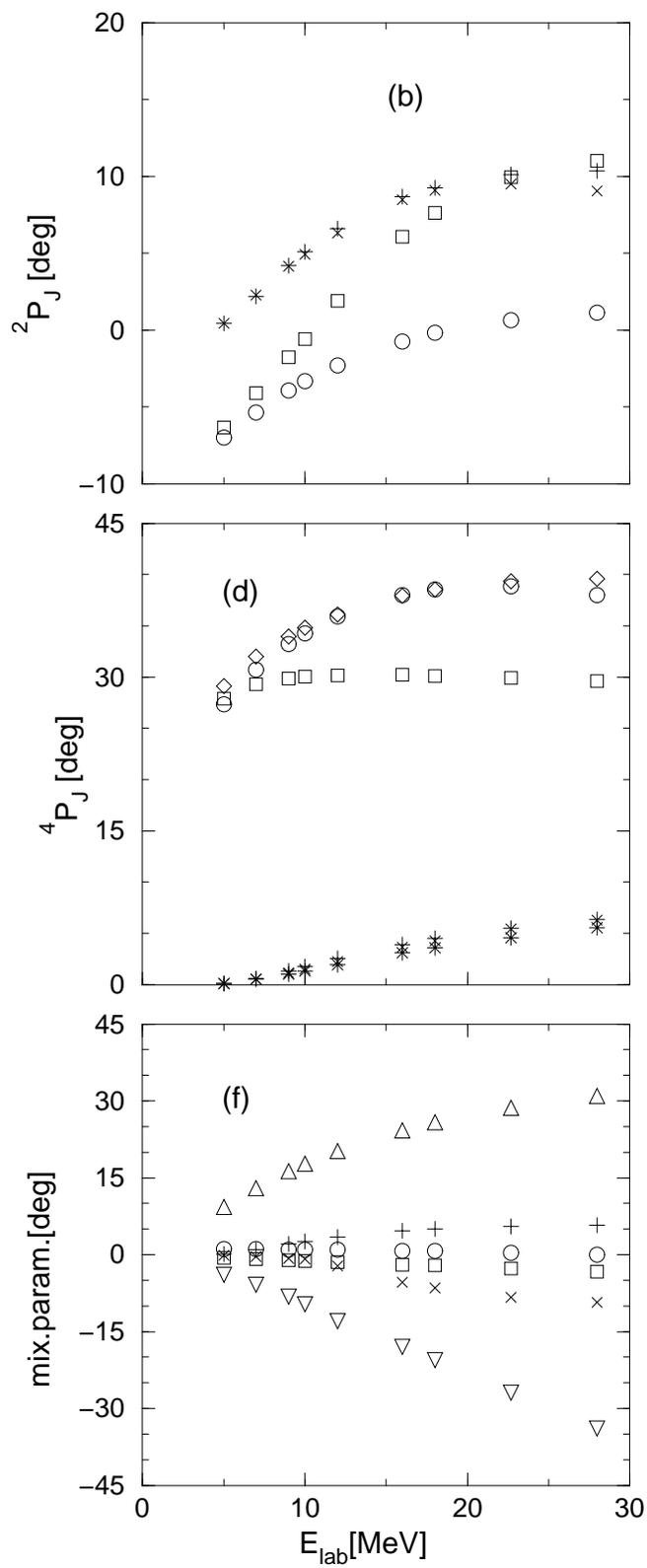
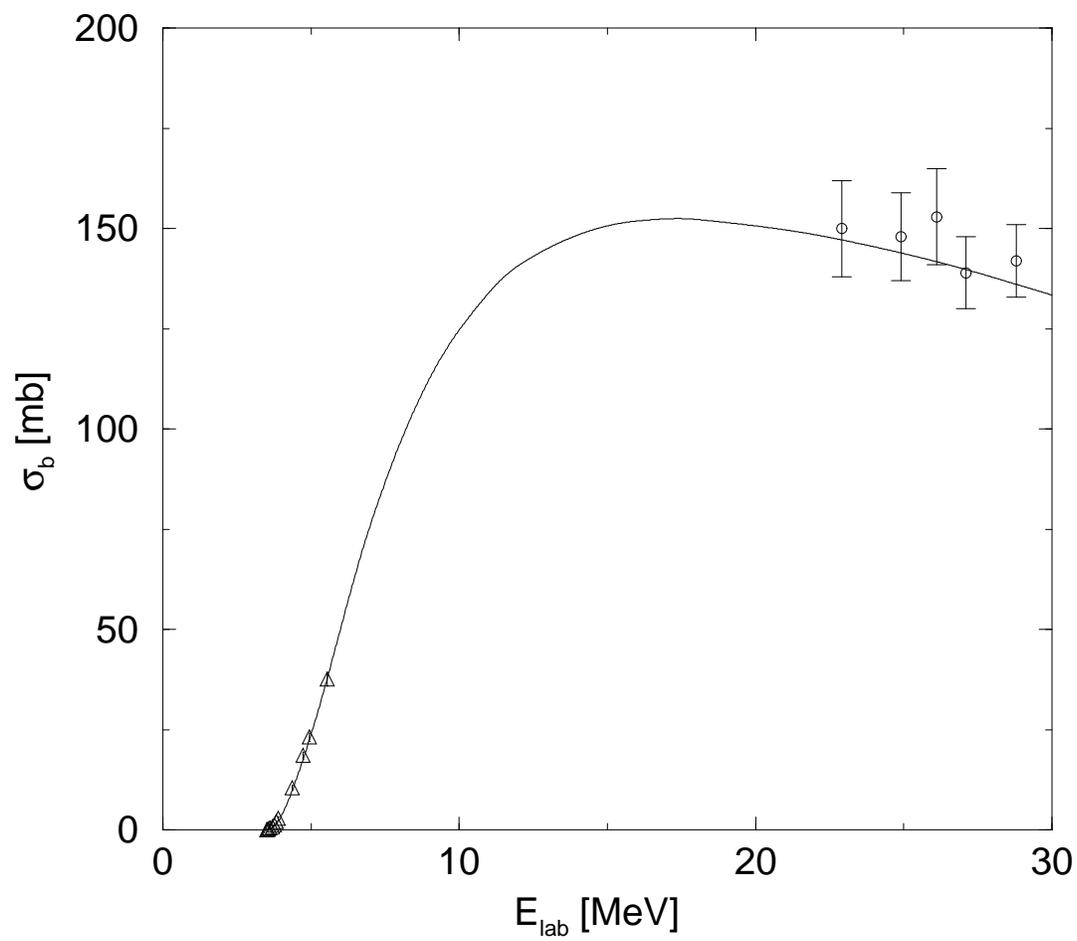
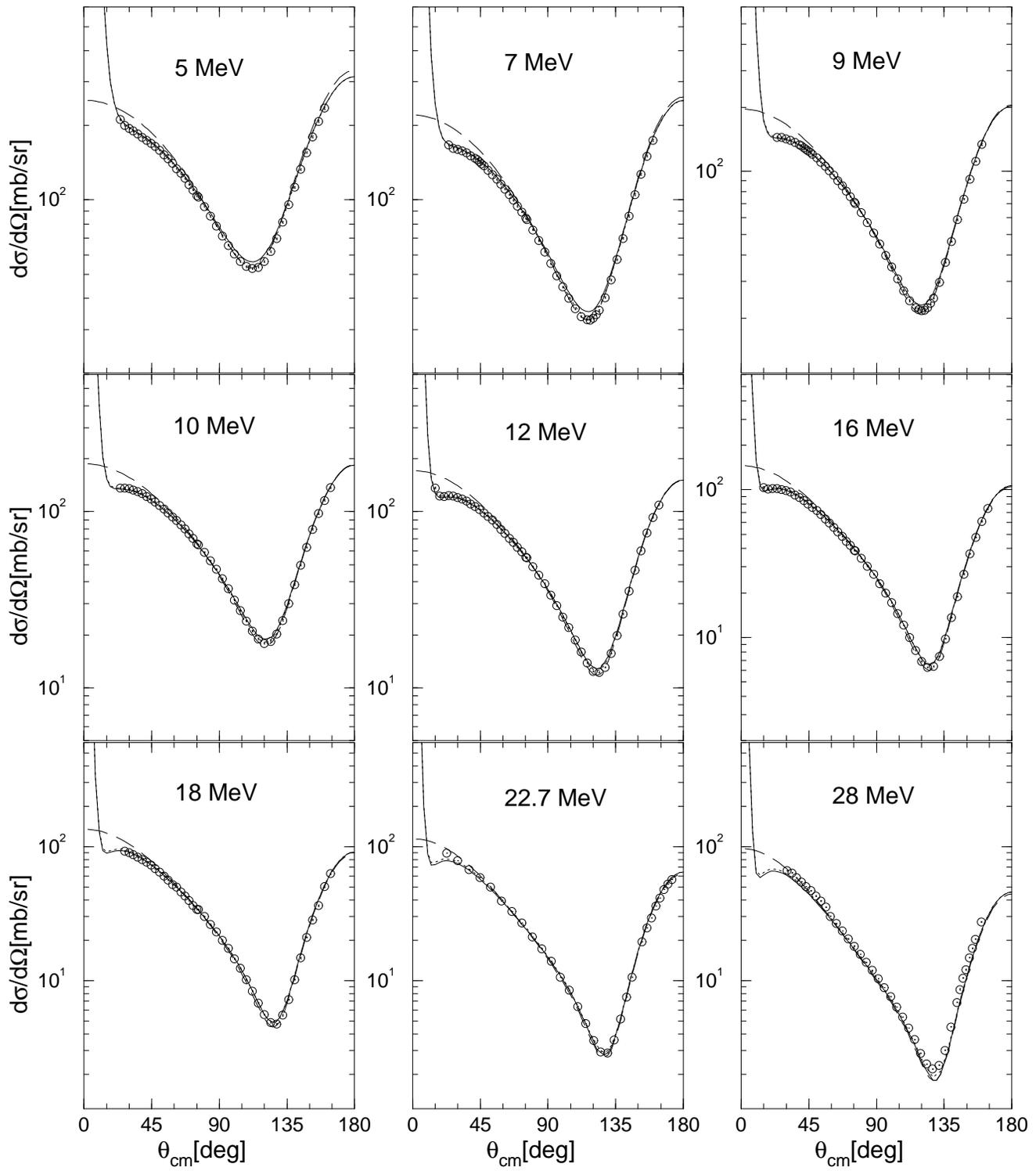
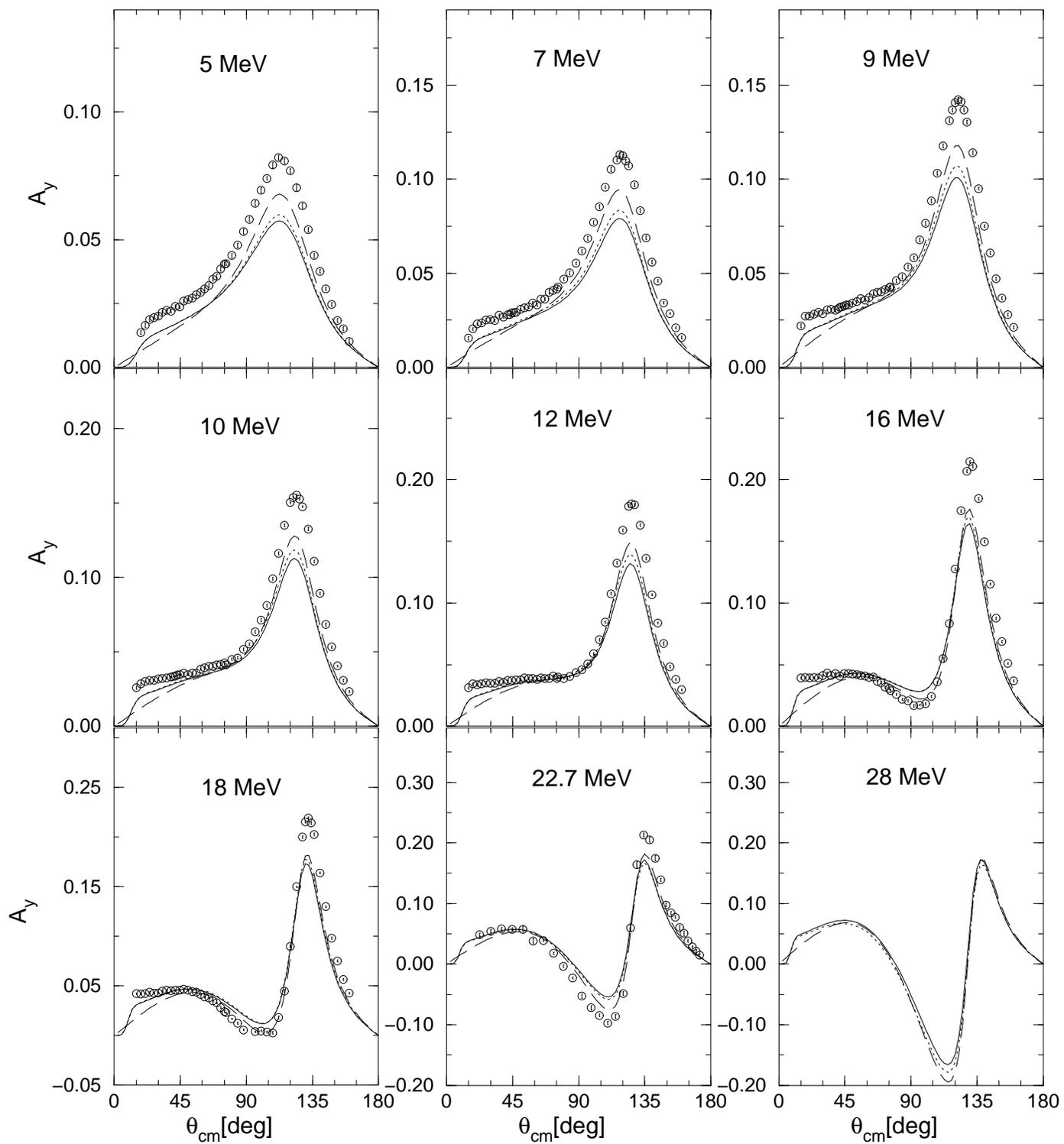
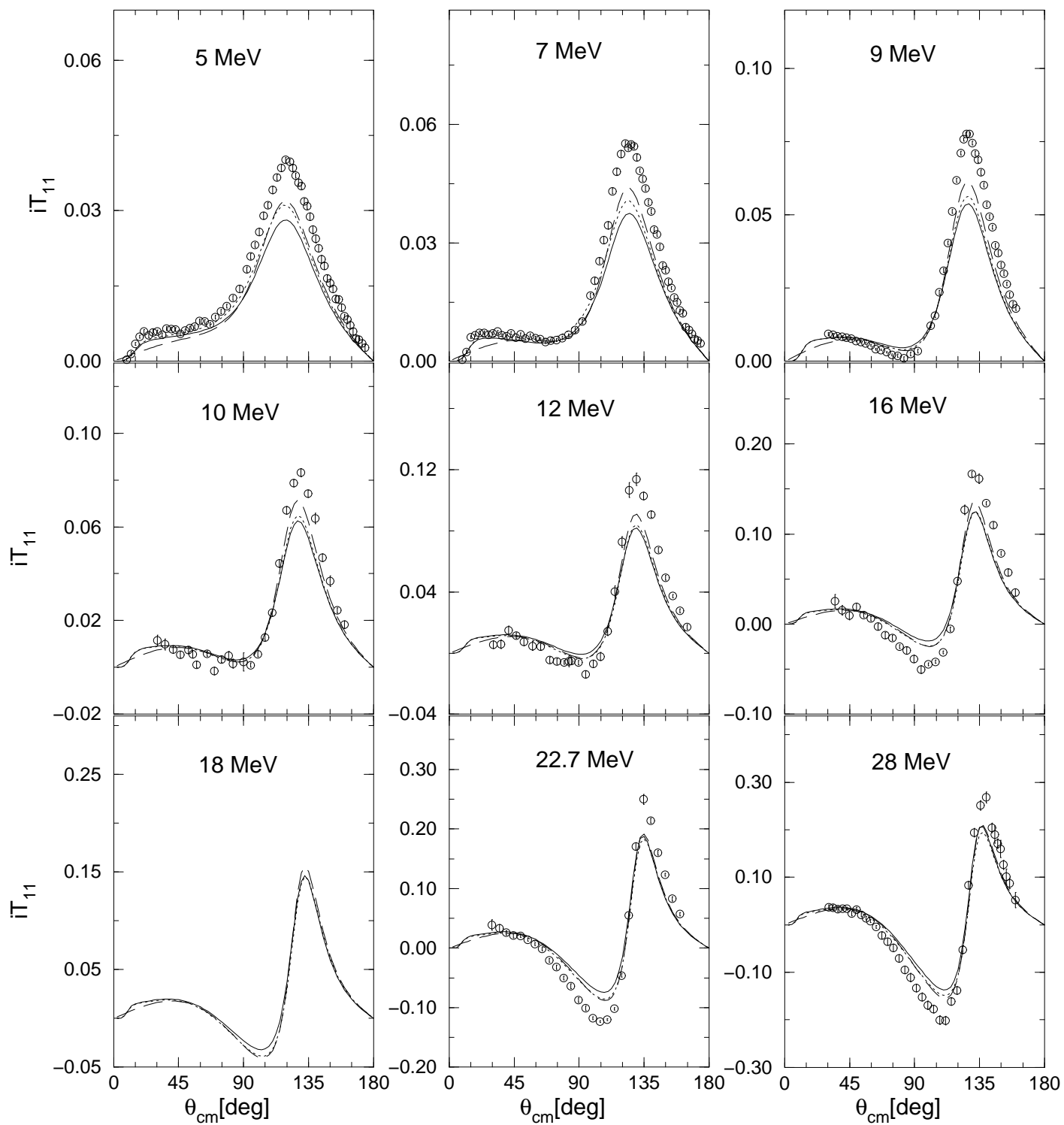
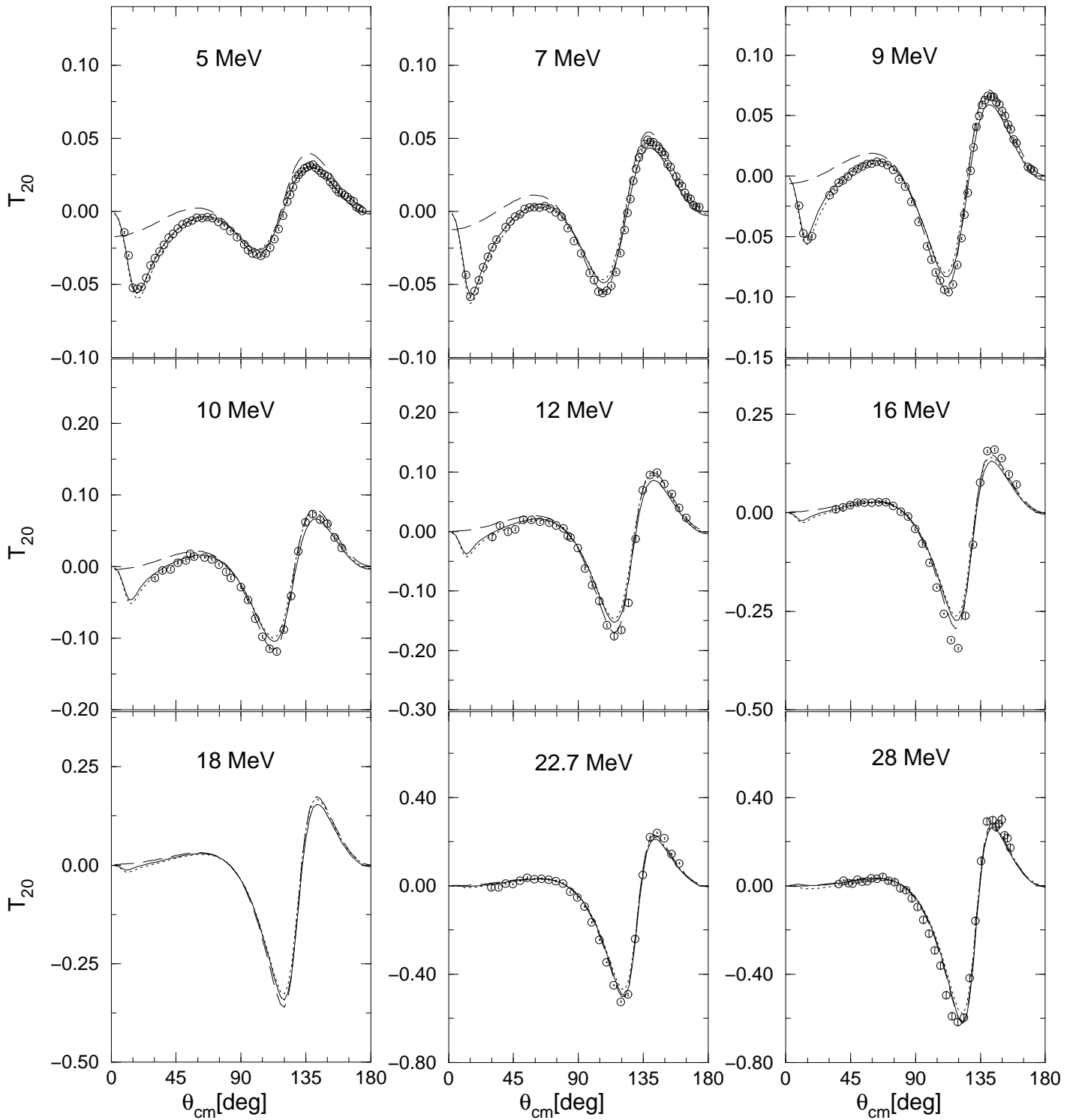
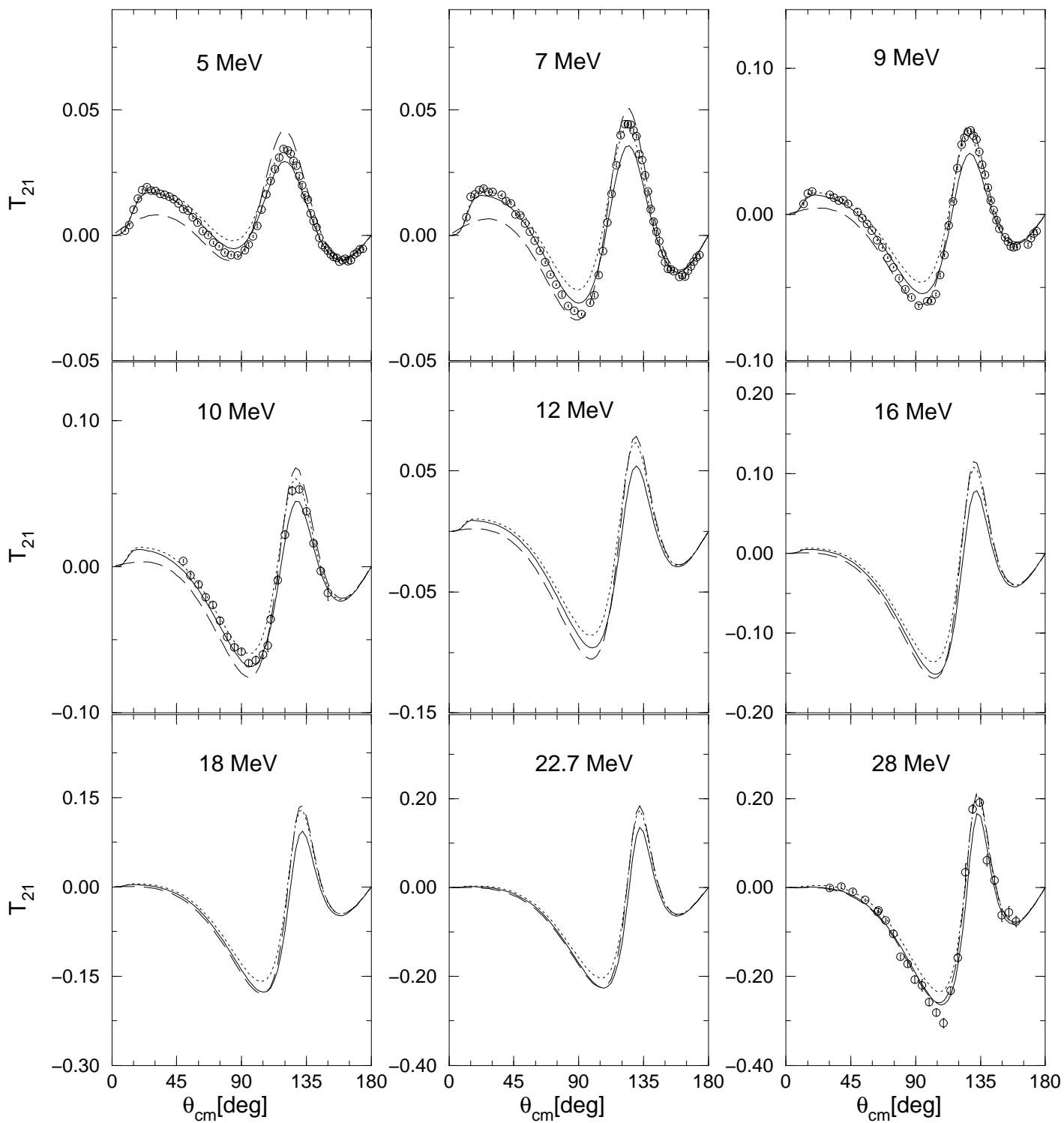
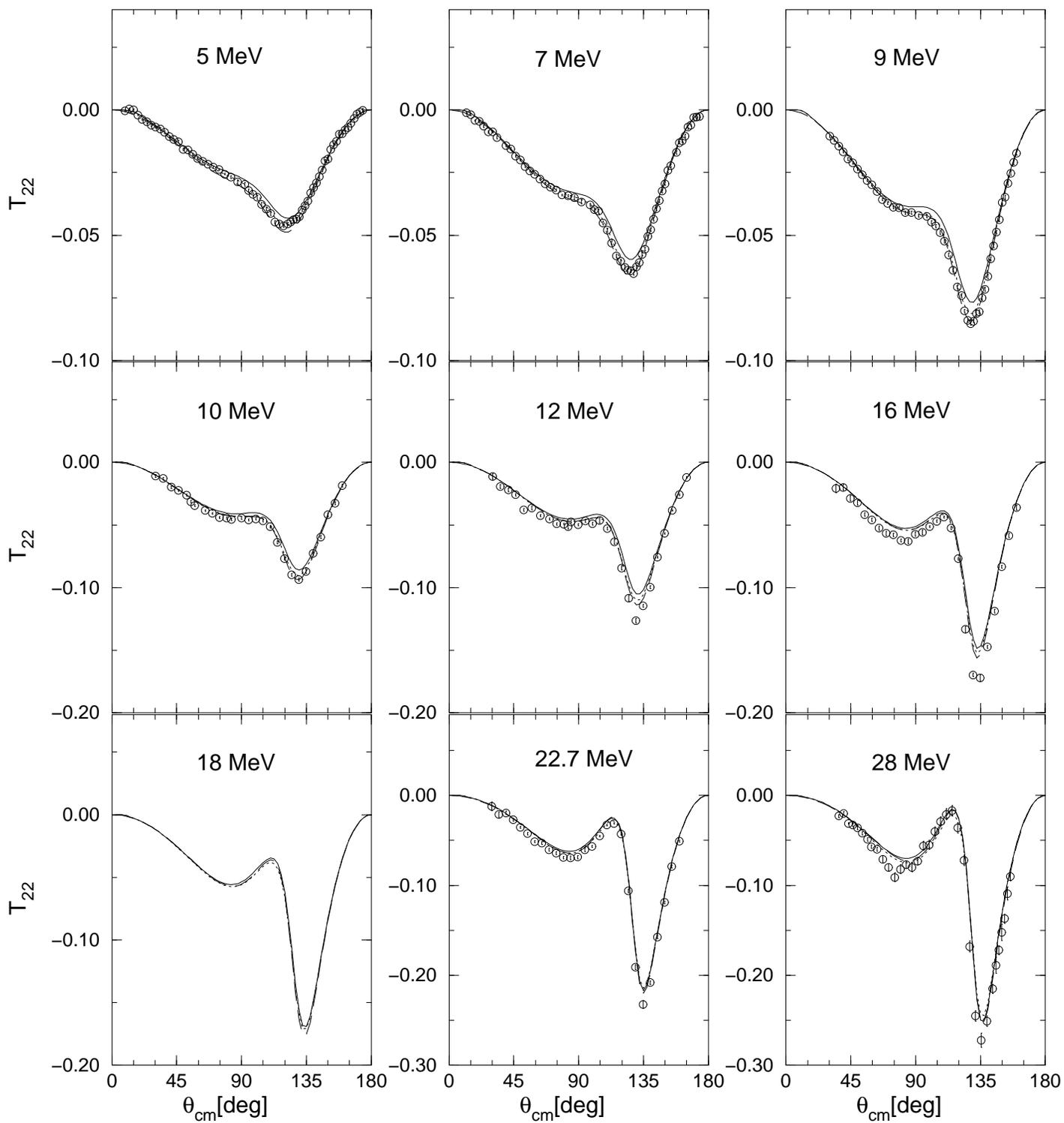
\end{document}